\documentclass{ws-ijmpa}
\usepackage[super,compress]{cite}
\usepackage{graphicx}
\usepackage{subfigure}
\usepackage{multirow}
\usepackage[bookmarksnumbered, pdfstartview=FitH,colorlinks,urlcolor=blue, citecolor=blue,linkcolor=blue,] {hyperref}
\begin{document}
\markboth{Han Miao and Jianyu Zhang}{Hyperon-Nucleus/Nucleon Scattering at BESIII}

%
\catchline{}{}{}{}{}
%

\title{Hyperon-Nucleus/Nucleon Scattering at BESIII}

\author{Han Miao \footnote{on behalf of BESIII Collaboration}}

\address{Institute of High Energy Physics, Beijing 100049, People's Republic of China\\
University of Chinese Academy of Sciences, Beijing 100049, People's Republic of China\\
miaohan@ihep.ac.cn}

\author{Jianyu Zhang}
\address{University of Chinese Academy of Sciences, Beijing 100049, People's Republic of China}

\maketitle

\begin{history}
\received{Day Month Year}
\revised{Day Month Year}
\end{history}

\begin{abstract}
	Utilizing the large quantity of hyperons and antihyperons produced by the decay of 10 billion $J/\psi$ and 2.7 billion $\psi(3686)$ collected at BESIII, the cross sections of several specific elastic or inelastic (anti-)hyperon-nucleus/nucleon rections have been measured via the scattering between the (anti-)hyperons and the nucleus in the dense objects of BESIII detector. The novel method developed in these works extends the research field and opens a new era for the experiments at $e^+ e^-$ colliders. The results of such measurements will definitely benefit a lot the precise probe of the (anti-)hyperon-nucleus/nucleon interactions and provide constraints for the studies of the potential of strong interaction, the origin of color confinement, the unified model for baryon-baryon interactions, and the internal structure of neutron stars. The desirable prospects of corresponding studies in the future Super Tau-Charm Factory (STCF) are also discussed in this report.

\end{abstract}

\ccode{PACS numbers: 13.75.Ev, 14.20.Jn}


\section{Why to measure}

Recently, the properties of hyperons in dense matter have attracted much interest due to their close connection with hypernuclei and the hyperon component in neutron stars~\cite{Tolos:2020aln}. Hyperons may exist within the inner layer of neutron stars whose structure strongly depends on the equation of state (EOS) of nuclear matter at supersaturation densities~\cite{Lattimer:2000nx}. The appearance of hyperons in the core softens the EOS, resulting in neutron stars with masses lower than 2$M_\odot$~\cite{Lonardoni:2014bwa}, where $M_\odot$ is the mass of the sun. However, studies based on observations from the LIGO and Virgo experiments~\cite{LIGOScientific:2018cki} indicate that the EOS can support neutron stars with masses above $1.97M_\odot$. This is the so-called ``hyperon puzzle in neutron stars'', warranting further experimental and theoretical studies.

The key point to solve the ``hyperon puzzle'' is to study the interactions between the baryons. To date, strong constraints and well-established models exist for nucleon-nucleon interactions~\cite{Vidana:2018bdi,Hiyama:2018lgs}, but there are still difficulties in precisely modeling hyperon-nucleon scattering, especially hyperon-hyperon interactions, due to the lack of experimental measurements. Until now, there have only been a few measurements for hyperon-nucleon scattering~\cite{Eisele:1971mk,Sechi-Zorn:1968mao,Alexander:1968acu,Kadyk:1971tc,Hauptman:1977hr,KEK-PSE-251:1997cno,KEK-PS-E289:2000ytt,Ahn:2005jz,J-PARCE40:2021qxa,J-PARCE40:2021bgw,CLAS:2021gur,J-PARCE40:2022nvq, BESIII:2023clq, BESIII:2023trh, BESIII:2024geh}, and only one for hyperon-hyperon scattering~\cite{ALICE:2022uso}, leaving theoretical models largely unconstrained~\cite{Haidenbauer:2005zh, Rijken:2010zzb, Polinder:2006zh,Polinder:2007mp, Haidenbauer:2013oca,Haidenbauer:2015zqb, Haidenbauer:2018gvg, Haidenbauer:2019boi, Ren:2019qow, Haidenbauer:2023qhf, Li:2016paq, Li:2016mln, Ishii:2006ec,Ishii:2012ssm, Beane:2006gf,Beane:2010em, Schaefer:2005fi, Fujiwara:2006yh}.

Experimental studies of hyperon-nucleon interactions still suffer a lot from the difficulty of obtaining a stable hyperon beam. Firstly, the lifetime of ground-state hyperons is usually of order $O(10^{-10})~{\rm s}$ due to the weak decay, which is too short to be a stable beam. Meanwhile, hyperons historically used for fixed-target experiments are commonly produced in the $K p$ and $\gamma p$ collisions, such as J-PARC and CLAS experiments, with a high hadronic background level. Compared with the fixed-target experiments, much more hyperons are accessible from the decay of charmonia produced at $e^+ e^-$ colliders, which have recently been used to measure the hyperon-nucleus interactions at BESIII~\cite{BESIII:2023clq, BESIII:2023trh, BESIII:2024geh}. Furthermore, abundant antihyperons produced in pair with hyperons bring exciting prospects to probe antihyperon-nucleus/nucleon interactions that have rarely been measured and studied~\cite{BESIII:2024geh}.

\section{How to measure}

\subsection{BESIII and BEPCII}

The BESIII detector is a magnetic spectrometer~\cite{BESIII:2009fln} located at the Beijing Electron Positron Collider (BEPCII). The cylindrical core of the BESIII detector consists of a helium-based  multilayer drift chamber (MDC), a plastic scintillator time-of-flight system (TOF), and a CsI(Tl) electromagnetic calorimeter (EMC), which are all enclosed in a superconducting solenoidal magnet providing a 1.0~T magnetic field.

In particularly, the BESIII detector has excellent performance on the reconstruction of long-lived particles, such as $K_S$ and ground-state hyperons ($\Lambda$, $\Sigma^{+,-}$, $\Xi^{0,-}$ and $\Omega^-$), and has published dozens of analyses involved with hyperon physics~\cite{BESIII:2022qax, BESIII:2021ypr, BESIII:2020fqg, BESIII:2018cnd, BESIII:2022lsz, BESIII:2023drj}.

\subsection{Method for the measurement}

\begin{figure}[htbp]
        \centering
        \includegraphics[width=0.8\textwidth]{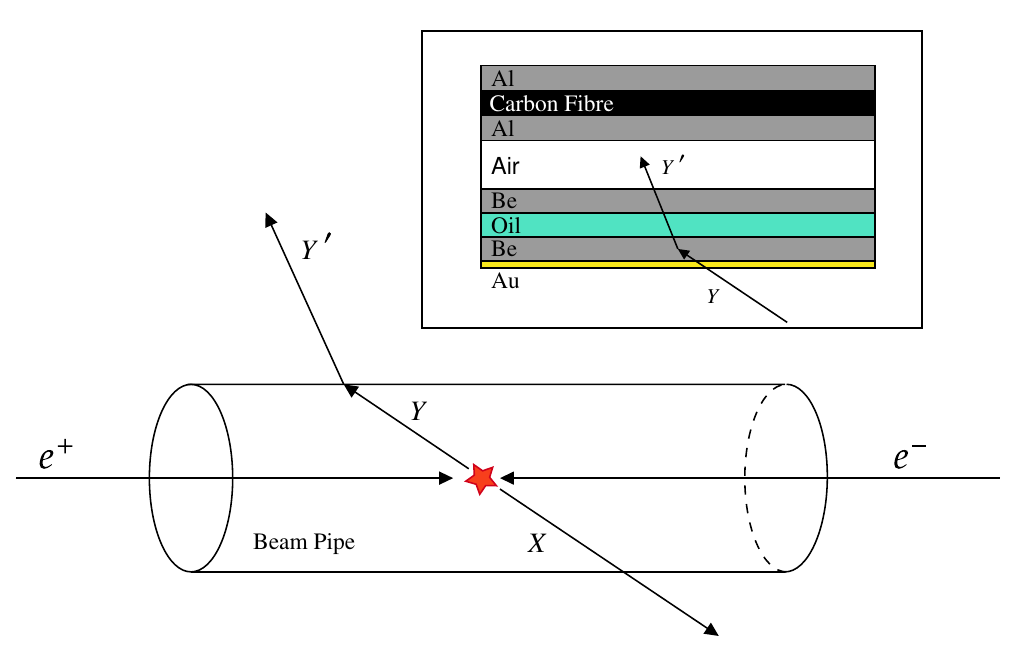}
        \caption{A schematic picture of hyperon-nucleon interactions at $e^+ e^-$ collider represented by BESIII. The symbol $Y$ denotes the hyperon interacting with the nucleus and $X$ represents the other particles produced in the $e^+ e^-$ collision together with $Y$. The symbol $Y^{\prime}$ denotes the particles produced in the hyperon-nucleus interaction. The structure of the target is shown in the top right corner.}
        \label{fig:r99bes}
\end{figure}

As shown in Fig.~\ref{fig:r99bes}, hyperon pairs or final states including hyperons are produced by the collision of $e^+ e^-$ inside the beam pipe and will fly in the direction of the momentum. Some of the hyperons are able to arrive at the beam pipe or the inner wall of MDC, which are the target of this work, before decaying and scattering elastically or inelastically with the nucleus inside the material. For an example at BESIII, $\Lambda$ hyperons may interact with the Be nucleus inside the beam pipe and subsequently converse into $\Sigma^+$ by exchanging a $\pi$ or $K$ meson with the nucleus. At the same time, Be nucleus will converse into another nuclide. The detailed information of the target can be found in Ref.~\cite{Dai:2022wpg}.

As shown in Fig.\ref{fig:r99bes}, considering the process $\psi \to XY,~YA \to Y^{\prime} A^{\prime}$ ($Y$ is the hyperon that interacts with nucleons, $X$ is the other particles except $Y$ produced in the decay of charmonia, $Y^{\prime}$ is the new hyperon created by the hyperon-nucleon interaction, $A$ and $A^{\prime}$ are the nucleus inside the target before and after the interaction), an experimental method called ``Double Tag" method can be used in such measurements.

The accurate information of the hyperon $Y$ that interacts with material, including momentum and direction, can be obtained by reconstructing the other particles denoted as $X$ among the final state, which is called ``single-tag". The yield of the hyperons of interest $N_{\rm ST}$ is obtained by fitting the recoil mass ($RM_{X}$) distribution of $X$, which is defined as
\begin{equation}
        RM_{X} = \sqrt{|p_{e^+ e^-} - p_{X}|^{2}},
\end{equation}
where $p_{e^+ e^-}$ and $p_{X}$ are the 4-momenta of the $e^+ e^-$ and $X$. $RM_{X}$ distribution will have a peak around the intrinsic mass of $Y$.

Then the hyperon produced by the hyperon-nucleon scattering denoted as $Y^{\prime}$ are reconstructed by the final states of its decay, which is called ``double-tag". The yield of $Y^{\prime}$ denoted as $N_{\rm DT}$ is obtained by fitting the distribution of the invariant mass of $Y^{\prime}$, which can be expressed as
\begin{equation}
        \label{equ:n_dt}
        N_{\rm DT}={\mathcal L}_{Y} \cdot \sigma(YA \rightarrow Y^{\prime}A^{\prime}) \cdot {\mathcal B}(Y^{\prime}) \cdot \epsilon_{\rm sig},
\end{equation}
where $\sigma(YA \rightarrow Y^{\prime}A^{\prime})$ is the cross section of the hyperon-nucleon process of interest, ${\mathcal B}(Y^{\prime})$ is the branching fraction of the decay channel used to reconstruct $Y^{\prime}$ and $\epsilon_{\rm sig}$ is the selection efficiency of $Y^{\prime}$ for the specific decay channel obtained from signal MC sample. To finally determine the cross section of such processes, a specially defined variable $\mathcal{L}_{Y}$, named ``effective luminosity'', is introduced to account for the properties of the target and the behavior of the incident hyperon beam and estimated using signal Monte Carlo (MC) samples described above. The detailed formulas to calculate effective luminosity can be found in Ref.~\cite{Dai:2022wpg}.

Using Eq.~(\ref{equ:n_dt}), the $\sigma(YA \rightarrow Y^{\prime}A^{\prime})$ is obtained to be
\begin{equation}
        \label{equ:cross_section}
        \sigma(YA \rightarrow Y^{\prime}A^{\prime}) = \frac{N_{\rm DT}}{\epsilon_{\rm sig} \cdot {\mathcal{L}_{Y}}} \cdot \frac{1}{{\mathcal B}(Y^{\prime})}.
\end{equation}

%
%
%
\section{What have been measured}

\subsection{$\Lambda N \to \Sigma^+ X$}

The reaction chain in this measurement is
\begin{equation}
	J/\psi \to \Lambda \bar{\Lambda},~\Lambda + N({\rm nucleus}) \to \Sigma^+ + X({\rm anything}),~\Sigma^+ \to p \pi^0,~\pi^0 \to \gamma \gamma.
\end{equation}

\begin{figure}[htbp]
		\centering
		\subfigure[$RM_{\bar{p} \pi^+}$ distribution in the single-tag side with the best fit overlaid.]{ \includegraphics[width=0.45\textwidth]{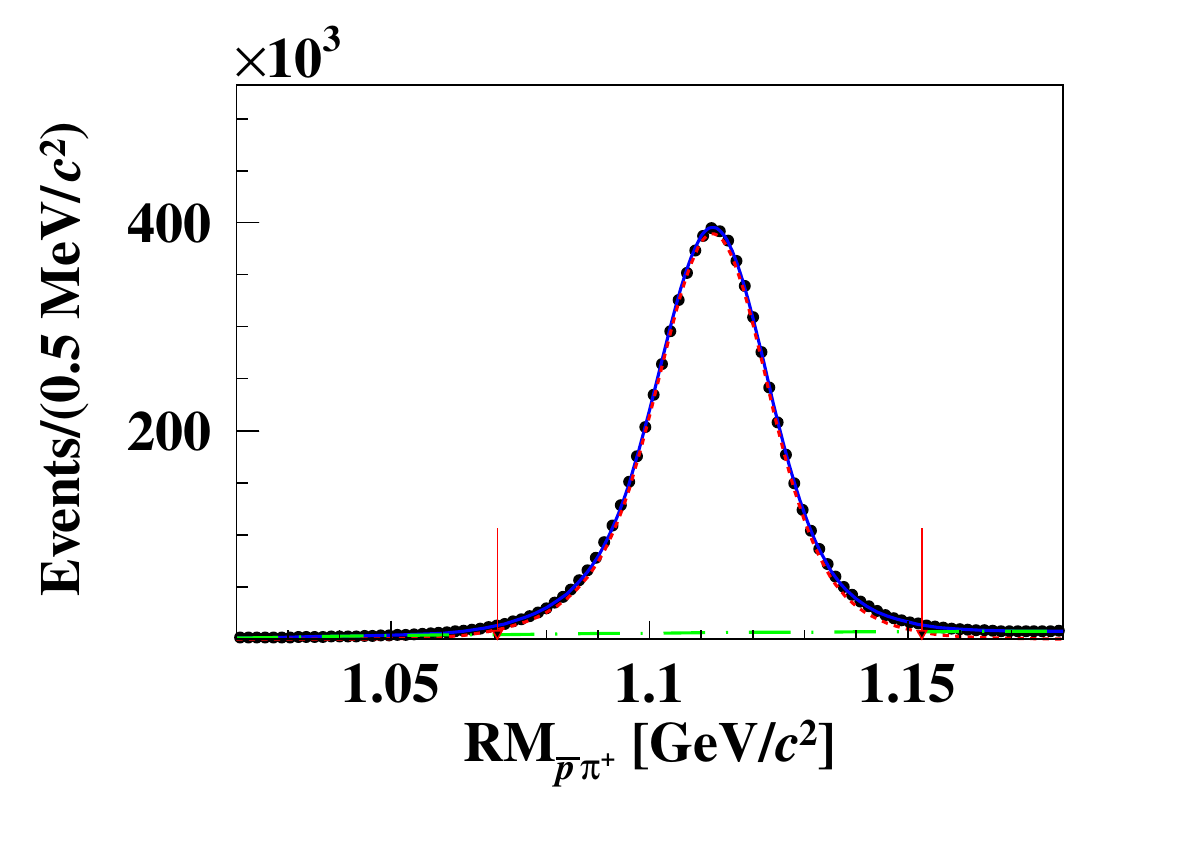} }
		\subfigure[$M_{p \pi^0}$ distribution in the double-tag side with the best fit overlaid.]{ \includegraphics[width=0.45\textwidth]{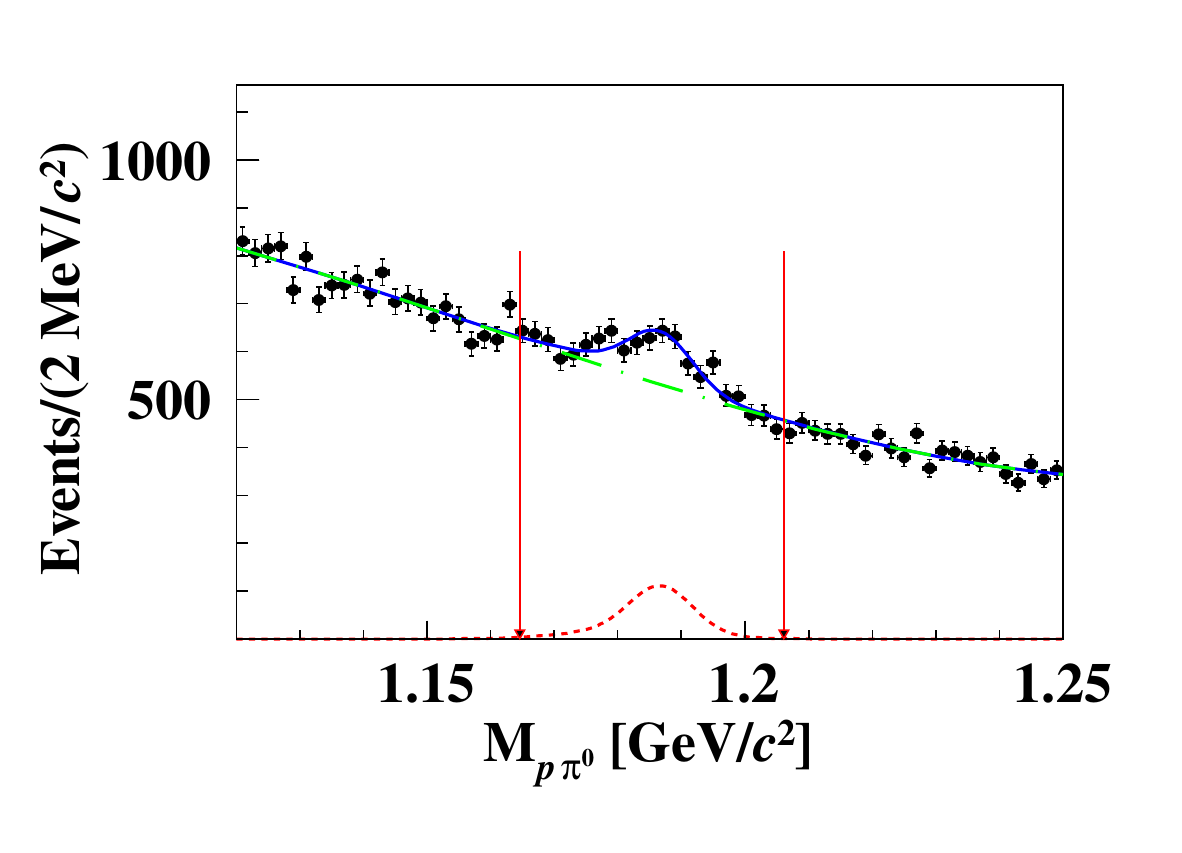} }
		\caption{}
		\label{fig:LambdaN_inelastic}
	\end{figure}

	Only $\bar{\Lambda}$ and $\Sigma^+$ are reconstructed in this work, without reconstructing other final states created in the hyperon-nucleon interactions. The momenta of the incident $\Lambda$ hyperons are about $p_{\Lambda} \approx 1.074~{\rm GeV}/c$, which is a small range about $0.017~{\rm GeV}/c$ above and below $1.074~{\rm GeV}/c$ due to the very small horizontal cossing angle of $e^+$ and $e^-$ beam. The crossing angle introduces a small boost of the whole system in the laboratory coordinate. The yields of the single-tagged events and double-tagged events are obtained by fitting to the recoil mass distribution of $\bar{p} \pi^+$ from $\bar{\Lambda}$ decay ($RM_{\bar{p} \pi^+}$) and the invariant mass distribution of $p \pi^0$ from $\Sigma^+$ decay ($M_{p \pi^0}$), respectively, as shown in Fig.~\ref{fig:LambdaN_inelastic}. Using the necessary parameters listed in Tab.~\ref{tab:LambdaN_inelastic} and incorperating the relative systematic uncertainty as 9.5\%, the cross section of $\Lambda + {^{9}{\rm Be}} \to \Sigma^+ + X$ is calculated to be $\sigma(\Lambda + {^{9}{\rm Be}} \to \Sigma^+ + X) = (37.3 \pm 4.7_{\rm stat.} \pm 3.5_{\rm syst.})~{\rm mb}$ at $p_{\Lambda} \approx 1.074~{\rm GeV}/c$. This is the first attempt to investigate $\Lambda$-nucleus interaction at an $e^+ e^-$ collider.

Taking the effective number of reaction protons in ${^{9}{\rm Be}}$ nucleus as 1.93, the cross section of $\Lambda p \to \Sigma^+ X$ for single proton is $\sigma(\Lambda p \to \Sigma^+ X) = (19.3 \pm 2.4_{\rm stat.} \pm 1.8_{\rm syst.})~{\rm mb}$. According to Ref.~\cite{Song:2021yab}, $\sigma(\Lambda p \to \Sigma^+ n)$ should be twice of $\sigma(\Lambda p \to \Sigma^0 p)$. The result from BESIII is consistent with the result in Ref.~\cite{Hauptman:1977hr} if neglecting the contribution of the reactions with three-body final states. More detailed information can be found in Ref.~\cite{BESIII:2023trh}.

\begin{table}[htbp]
        \centering
        \caption{Inputs used to calculate the cross section of $\Lambda + ^{9}{\rm Be} \rightarrow \Sigma^+ + X$.}
        \begin{tabular}{cc}
                \hline
                \hline
                Parameter               &               Value           \\
                \hline
                $N_{\rm DT}$            &               $795\pm101$             \\
                $\epsilon_{\rm sig}$    &               24.32\%                 \\
                $\mathcal{L}_{\Lambda}$ &               $(17.00\pm0.01)\times10^{28}~{\rm cm^{-2}}$     \\
                $\mathcal{B}(\Sigma^+ \to p \pi^0)$     &       $(51.57\pm0.30)\%$      \\
                \hline
                \hline
        \end{tabular}
        \label{tab:LambdaN_inelastic}
\end{table}

\subsection{$\Xi^0 n \to \Xi^- p$}
	\label{sec:Xi_inelastic}

The reaction chain in this measurement is
\begin{equation}
	J/\psi \to \Xi^0 \bar{\Xi}^0,~\bar{\Xi}^0 \to \bar{\Lambda} \pi^0,~\bar{\Lambda} \to \bar{p} \pi^+,~\pi^0 \to \gamma \gamma,~\Xi^0 n \to \Xi^- p,~\Xi^- \to \Lambda \pi^-,~\Lambda \to p \pi^-.
\end{equation}

\begin{figure}[htbp]
		\centering
		\subfigure[Signal yield of $\Xi^0 n \to \Xi^- p$.]{ \label{fig:subfig:Xi_fit} \includegraphics[width=0.45\textwidth]{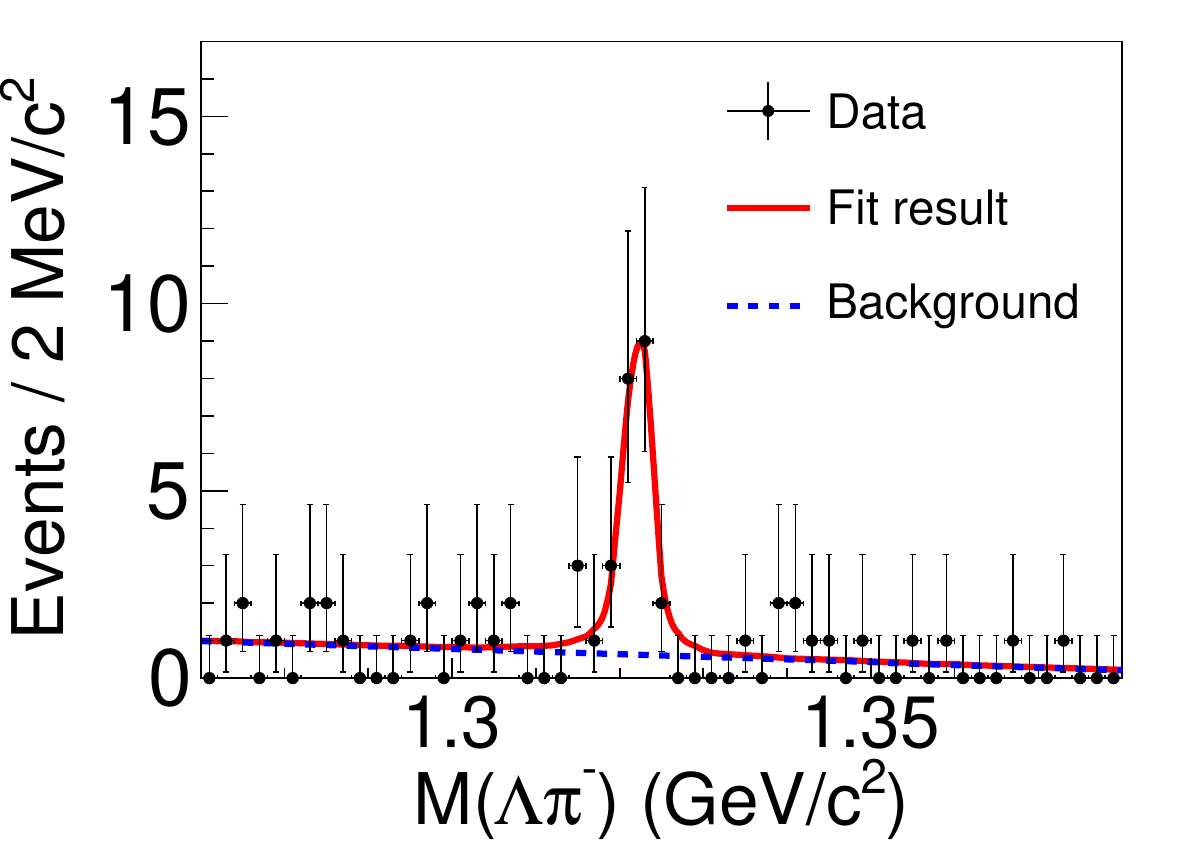} }
		\subfigure[$M_{\Lambda \pi^-}$ versus $R_{xy}$.]{ \label{fig:subfig:mXi_vs_Rxy} \includegraphics[width=0.45\textwidth]{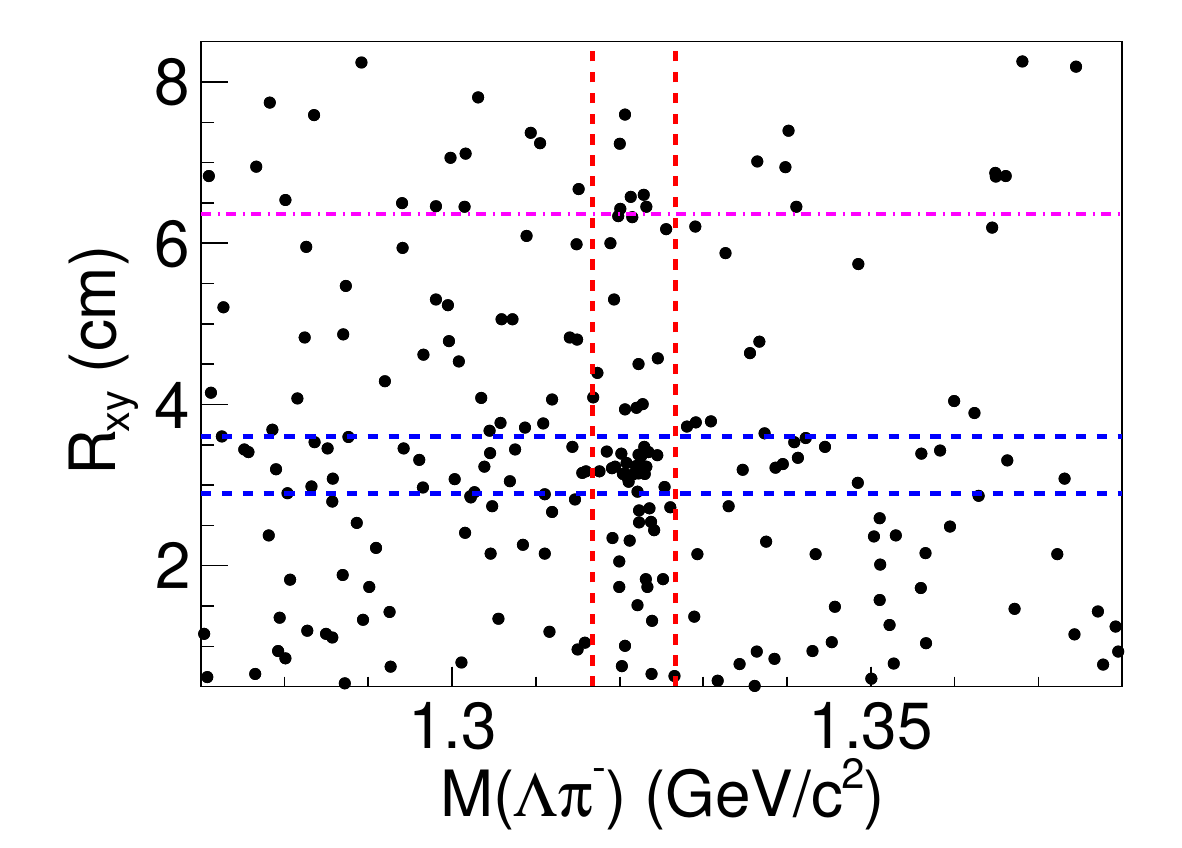} }
		\caption{}
	\end{figure}

	All the final particles in the reaction are reconstructed. The momenta of the incident $\Lambda$ hyperons are about $p_{\Lambda} \approx 0.818~{\rm GeV}/c$. The signal yield is extracted by fitting the invariant distribution of $\Lambda \pi^-$ to be $N_{\rm sig} = 22.9 \pm 5.5$ with a significance of $7.1\sigma$, as shown in Fig.~\ref{fig:subfig:Xi_fit}. The two-dimensional distribution of $M_{\Lambda \pi^-}$ versus the distance between the scattering point and the beam ($R_{xy}$) is also displayed in Fig.~\ref{fig:subfig:mXi_vs_Rxy}. $R_{xy}$ is obtained by performing a vertex fit to the reconstructed $\Lambda$ and $\pi^-$. Two obvious enhancements can be seen around the positions of the beam pipe and inner wall of the MDC.

Since the beam pipe is composed of layers of composite material, as shown in Fig.~\ref{fig:r99bes}, the cross section of the reaction between $\Xi^0$ baryons and $^{9}\rm{Be}$ nuclei $\sigma(\Xi^0+{^9\rm{Be}}\to\Xi^-+p+{^8\rm{Be}})$ is extracted using
\begin{equation}
    \sigma(\Xi^0+{^9\rm{Be}}\to\Xi^-+p+{^8\rm{Be}}) = \frac{N^{\rm{sig}}} {\epsilon \mathcal{B} \mathcal{L}_{\rm{eff}} },
\end{equation}
where $\epsilon$ is the selection efficiency, $\mathcal{B}$ is the product of the branching ratios of all intermediate resonances, defined as $\mathcal{B}\equiv\mathcal{B}(\bar{\Xi}^0\to\bar{\Lambda}\pi^0)\mathcal{B}(\bar{\Lambda}\to\bar{p}\pi^+)\mathcal{B}(\pi^0\to\gamma\gamma)
\mathcal{B}(\Xi^-\to\Lambda\pi^-)\mathcal{B}(\Lambda\to p\pi^-)$, and $\mathcal{L}_{\rm{eff}}$ is the effective luminosity of the $\Xi^0$ flux produced from $J/\psi\to\Xi^0\bar{\Xi}^0$ and the distribution of target materials, as shown in the following formula:
\begin{equation}
    \mathcal{L}_{\rm{eff}} = \frac{\it{N}_{\it{J}/\psi}\mathcal{B}_{\it{J}/\psi}}{\rm{2}+\frac{\rm{2}}{\rm{3}}\alpha} \int_{a}^{b}\int_{\rm{0}}^{\pi} (\rm{1}+\alpha \rm{cos}^2\theta) \it{e}^{-\frac{x}{\rm{sin}\theta \it{\beta\gamma L}}}N(x)C(x) \rm{d}\theta \rm{d}\it{x}.
\end{equation}
The detailed definition and the values of the symbols in the formulae above can be found in Ref.~\cite{BESIII:2023clq}. Finally, the cross section of $\Xi^0+{^9\rm{Be}}\to\Xi^-+p+{^8\rm{Be}}$ is calculated to be $\sigma(\Xi^0+{^9\rm{Be}}\to\Xi^-+p+{^8\rm{Be}}) = (22.1 \pm 5.3_{\rm stat.} \pm 4.5_{\rm syst.})~{\rm mb}$ at $p_{\Xi^0} \approx 0.818~{\rm GeV}/c$. Taking the effective number of the reaction neutrons in ${^{9}{\rm Be}}$ nucleus as 3, the $\Xi^0-n$ cross section is calculated to be $\sigma(\Xi^0 n \to \Xi^- p) = (7.4 \pm 1.8_{\rm stat.} \pm 1.5_{\rm syst.})~{\rm mb}$, which is consistent with theoretical predictions~\cite{Polinder:2007mp, Haidenbauer:2015zqb, Haidenbauer:2018gvg}. In addition, $H$-dibaryon is also searched for in this work and no significant signals are seen.

This work is the first study of hyperon-nucleon interactions at $e^+ e^-$ colliders, openning up a new direction for such research. More detailed information can be found in Ref.~\cite{BESIII:2023clq}.

\subsection{$\Lambda (\bar{\Lambda}) p \to \Lambda (\bar{\Lambda}) p$}

The reaction chain in this measurement is
\begin{equation}
	J/\psi \to \Lambda \bar{\Lambda},~\Lambda p \to \Lambda p,~\Lambda \to p \pi^-,~\bar{\Lambda} \to \bar{p} \pi^+.
\end{equation}

\begin{figure}[htbp]
	\centering
	\subfigure[$M_{\Lambda p}$ distribution with the best fit overlaid.]{ \includegraphics[width=0.45\textwidth]{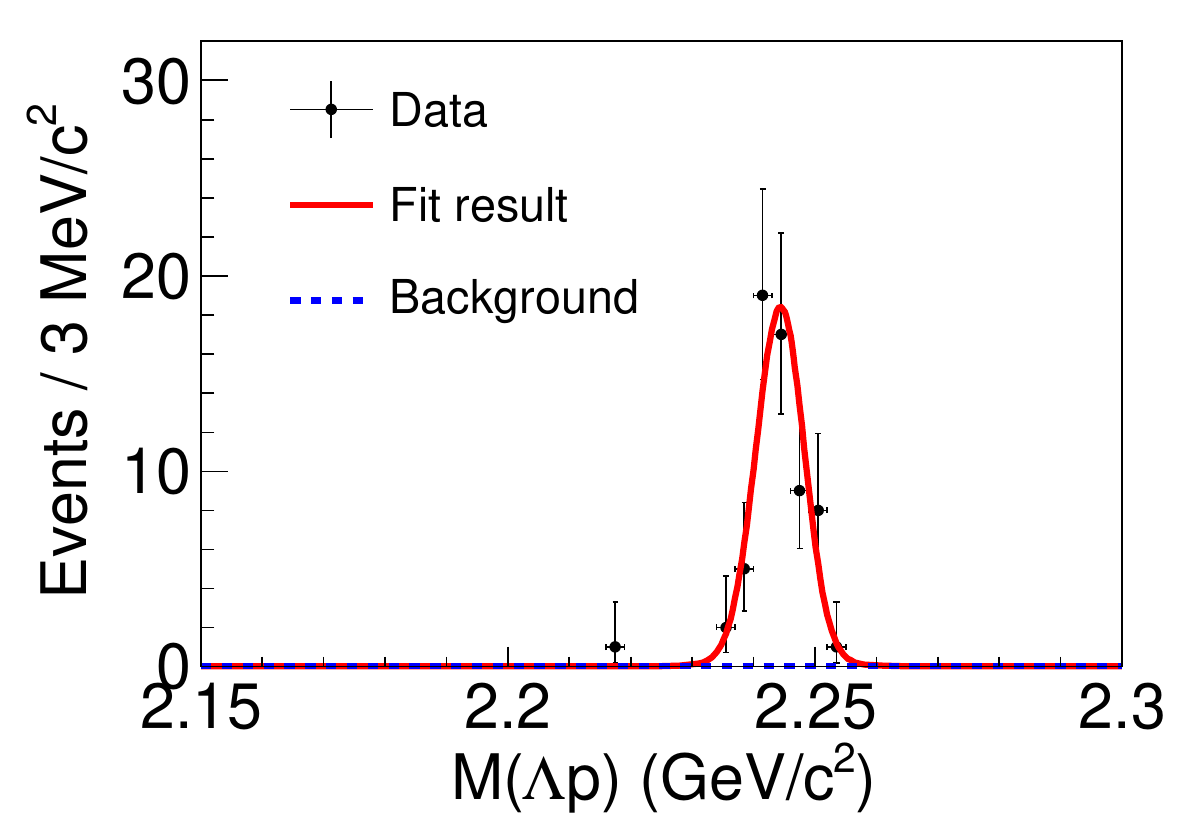} }
	\subfigure[$M_{\bar{\Lambda} p}$ distribution with the best fit overlaid.]{ \includegraphics[width=0.45\textwidth]{fit_mXia1.pdf} }
	\caption{}
	\label{fig:Lambda_elastic}
\end{figure}

Considering only the interaction between the incident $\Lambda (\bar{\Lambda})$ and hydrogen nucleus, the invariant mass of the reconstructed $\Lambda (\bar{\Lambda}) p$ will form a peak around $2.243~{\rm GeV}/c^{2}$ within a range of $\pm 0.005~{\rm GeV}/c^{2}$. Since the hydrogen nucleus can be treated as a static proton at the energy point cared about, the absolute yield of the $\Lambda- p$ elastic scattering instead of the $\Lambda$-nucleus interaction can be obtained by fitting the $M_{\Lambda (\bar{\Lambda}) p}$ distribution as shown in Fig.~\ref{fig:Lambda_elastic}.

Clear enhancements are seen around $2.243~{\rm GeV}/c^{2}$, corresponding to the reactions $\Lambda (\bar{\Lambda}) p \to \Lambda (\bar{\Lambda}) p$. Using the same method mentioned in Sec.~\ref{sec:Xi_inelastic}, the cross sections of the two reactions are determined to be $\sigma(\Lambda p \to \Lambda p) = (12.2 \pm 1.6_{\rm stat.} \pm 1.1_{\rm syst.})~{\rm mb}$ and $\sigma(\bar{\Lambda} p \to \bar{\Lambda} p) = (17.5 \pm 2.1_{\rm stat.} \pm 1.6_{\rm syst.})~{\rm mb}$ at $p_{\Lambda(\bar{\Lambda})} \approx 1.074~{\rm GeV}/c$ within $-0.9 < {\rm cos}\theta < 0.9$. The result is consistent with the result reported by CLAS collaboration~\cite{CLAS:2021gur}.

\begin{figure}[htbp]
	\centering
	\subfigure[Differential cross section with respect to ${\rm cos}\theta_{\Lambda}$ for $\Lambda p \to \Lambda p$ reaction.]{ \includegraphics[width=0.45\textwidth]{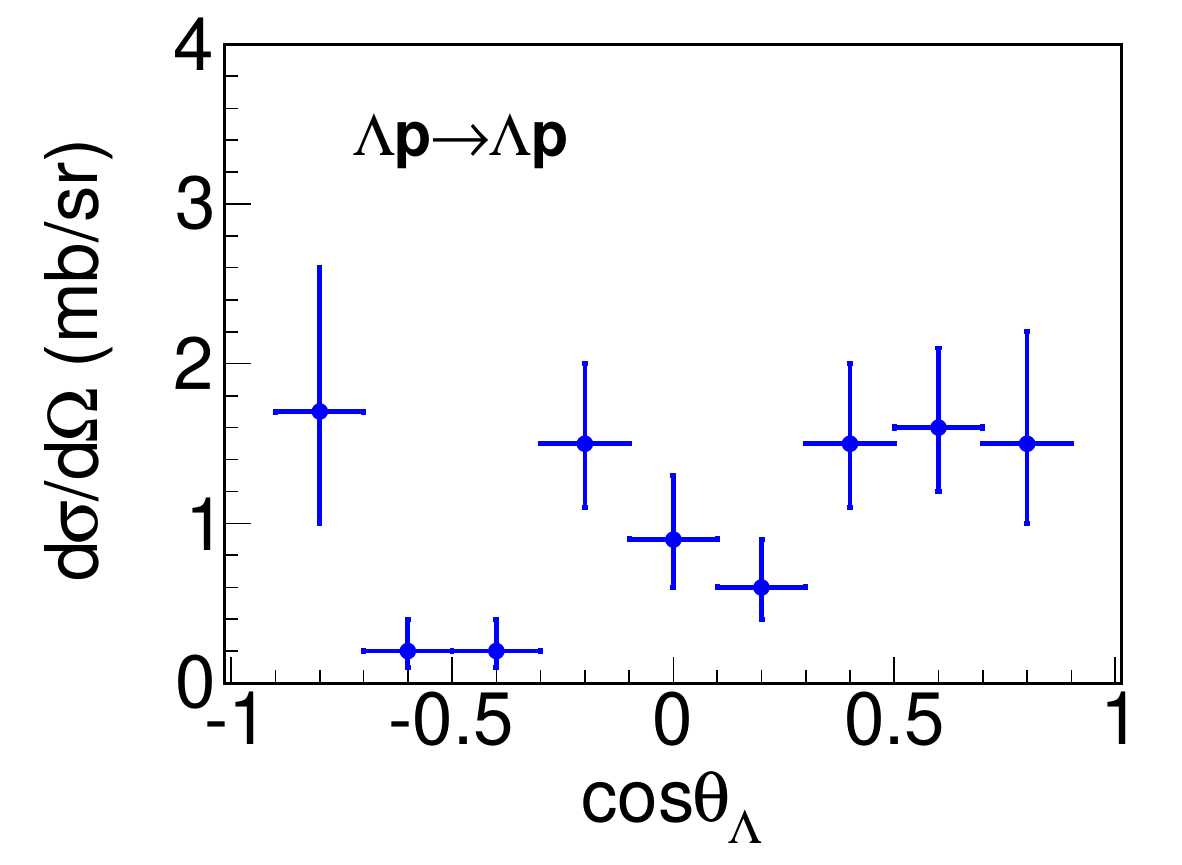} }
	\subfigure[Differential cross section with respect to ${\rm cos}\theta_{\bar{\Lambda}}$ for $\bar{\Lambda} p \to \bar{\Lambda} p$ reaction.]{ \includegraphics[width=0.45\textwidth]{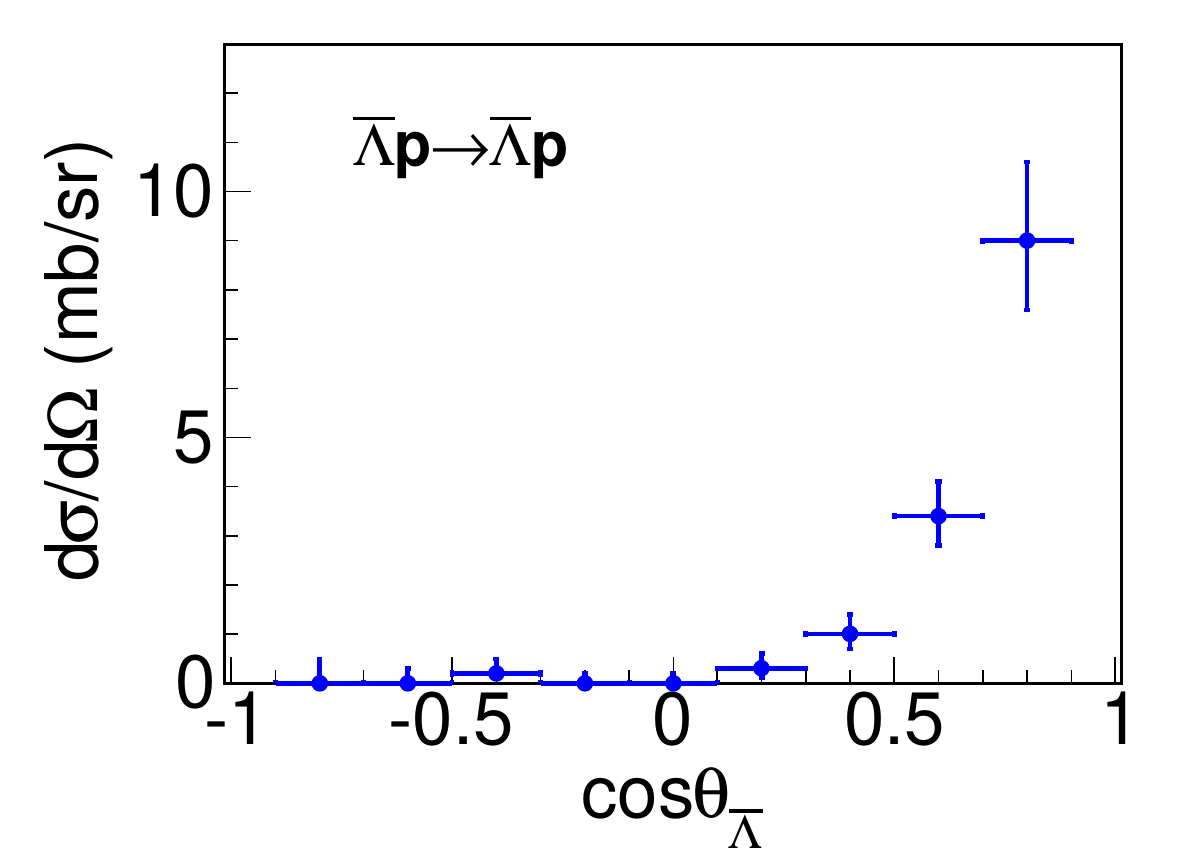} }
	\caption{}
	\label{fig:diff_cross_section}
\end{figure}

The differential cross sections of the two reactions are measured within $-0.9 < {\rm cos}\theta < 0.9$, as shown in Fig.~\ref{fig:diff_cross_section}, while there is a slight tendency of forward scattering for $\Lambda p \to \Lambda p$, and a strong forward peak for $\bar{\Lambda} p \to \bar{\Lambda} p$.

This work is the first measurement of antihyperon-nucleon scattering. More detailed information can be found in Ref.~\cite{BESIII:2024geh}.

\section{What will be measured}

\subsection{Potential channels at BESIII}

Considering the statistics of the dataset collected by BESIII now, there are several other channels of hyperon-nucleon interactions can be studied, which has been partially listed in Tab.~\ref{tab:potential_channel}. More results from BESIII collaboration will come out soon.

\begin{table}[htbp]
       \centering
       \caption{Potential channels of hyperon-nucleon interactions that can be measured at BESIII.}
       \begin{tabular}{cc}
               \hline
               \hline
               Hyperon Beam                                    &               Scattering Process                                      \\
               \hline
	       $J/\psi(\psi(3686)) \to p K^{-} \bar{\Lambda} + c.c.$   &       \multirow{2}*{$\Lambda p \to \Sigma^+ n/K^+ \Lambda \Lambda / K^+ \Xi^- p$, $\Lambda n \to \Sigma^- p$}	\\
	       $J/\psi(\psi(3686)) \to \Lambda \bar{\Lambda}$		&	~				\\
	       \hline
	       $J/\psi(\psi(3686)) \to \Sigma^- \bar{\Sigma}^+$		&	$\Sigma^- p \to \Sigma^0 n/\Lambda n$			\\
	       \hline
	       $J/\psi(\psi(3686)) \to \Sigma^+ \bar{\Sigma}^-$		&	$\Sigma^+ n \to \Sigma^0 p / \Lambda p$			\\
	       \hline
	       $J/\psi(\psi(3686)) \to \Xi^- \bar{\Xi}^+$		&	$\Xi^- p \to \Xi^0 n/\Lambda \Lambda$			\\
	       \hline
	       $J/\psi(\psi(3686)) \to \Xi^0 \bar{\Xi}^0$		&	$\Xi^0 n \to \Xi^- p/\Lambda \Lambda$			\\
	       \hline
	       \hline
       \end{tabular}
	\label{tab:potential_channel}
\end{table}

\subsection{Prospects at STCF}

In the near future, new generation of $e^+ e^-$ collider in proposal with a super high intensity beam called Super Tau-Charm Facility (STCF) will reach a peak luminosity of $1\times10^{35}~{\rm cm^{-2}s^{-1}}$~\cite{Achasov:2023gey}, which will be 100 times of the present BEPCII. Especially, the uncertainty of center-of-mass energy will improve from $1.2~{\rm MeV}$ to $20 \sim 80~{\rm keV}$ if monochromatic collision can be realized at the narrow resonances such as $J/\psi$~\cite{Telnov:2020rxp}, so that the cross section of $J/\psi$ production will further increase by 10 times. Given that there will be $10^{12} \sim 10^{13}$ $J/\psi$ produced at STCF per year, $10^6 \sim 10^7$ scattering events will be available. More precise measurements of the cross sections can be performed and the measurements of the interactions between $\Omega^-$ and nucleon become possible. Furthermore, the differential cross sections of the hyperon-nucleon interactions will be presented and the momentum-dependent cross sections can also be obtained using the three-body hyperon-involved decays of charmonia. In addition, potential hypernuclei can also be searched for at STCF.

Especially, the polarization of the hyperon pairs from the decay of charmonia produced at BESIII has been studied thoroughly from both experimental~\cite{BESIII:2022qax, BESIII:2021ypr, BESIII:2020fqg, BESIII:2018cnd, BESIII:2022lsz, BESIII:2023drj} and theoretical~\cite{Faldt:2017kgy} sides in recent years. It is found that the polarization will be a function of the polar angle of the hyperons, which inspires us to suppose that the relationship between the cross sections and the polarization of the incident hyperons can also be studied in detail by constraining the phase space. The polarization-dependent mechanism for such processes will be of great significance for figuring out the role of spin in hyperon-nucleon interactions and the potential of strong interaction. Meanwhile, large polarization effects have been observed in the produced hadrons in the study of elastic hadronic scattering. Similarly, the polarization of the produced hyperons in hyperon-nucleon interactions will also be a possible topic for the research at BESIII by measuring the angular distribution of the particles from the hyperon decay, which will be potential clues for the spin-involved interaction between hyperon and nucleon.

All of these exciting, attractive measurements will significantly proceed the research on the hyperon-nucleus/nucleon interactions and help a lot in revealing the puzzle of the internal structure of neutron stars.

\section{Summary}

A novel method has been developed for the measurement of hyperon-nucleus/nucleon interactions at BESIII and three results have been reported up to now. The cross section of $\Xi^0 n \to \Xi^- p$ is firstly measured with $\Xi^0$ beam from the decay $J/\psi \to \Xi^0 \bar{\Xi}^0$ based on 10 billion $J/\psi$ data to be $\sigma(\Xi^0+{^9\rm{Be}}\to\Xi^-+p+{^8\rm{Be}}) = (22.1 \pm 5.3_{\rm stat.} \pm 4.5_{\rm syst.})~{\rm mb}$ at $p_{\Xi^0} \approx 0.818~{\rm GeV}/c$, which is the first study of hyperon-nucleon interaction in $e^+ e^-$ colliders, openning a new direction for such research. The cross section of $\Lambda + {^{9}{\rm Be}} \to \Sigma^+ + X$ is studied wth $\Lambda$ from $J/\psi \to \Lambda \bar{\Lambda}$ to be $\sigma(\Lambda + {^{9}{\rm Be}} \to \Sigma^+ + X) = (37.3 \pm 4.7_{\rm stat.} \pm 3.5_{\rm syst.})~{\rm mb}$ at $p_{\Lambda} \approx 1.074~{\rm GeV}/c$, which is the first attempt to investigate $\Lambda$-nucleus interaction at $e^+ e^-$ colliders. Furthermore, the cross sections of elastic scatterings between $\Lambda (\bar{\Lambda})$ and proton are also measured to be $\sigma(\Lambda p \to \Lambda p) = (12.2 \pm 1.6_{\rm stat.} \pm 1.1_{\rm syst.})~{\rm mb}$ and $\sigma(\bar{\Lambda} p \to \bar{\Lambda} p) = (17.5 \pm 2.1_{\rm stat.} \pm 1.6_{\rm syst.})~{\rm mb}$ at $p_{\Lambda(\bar{\Lambda})} \approx 1.074~{\rm GeV}/c$ within $-0.9 < {\rm cos}\theta < 0.9$, which is the first study of the antihyperon-nucleon interactions. The differential cross sections are also presented in this work. With more statistics in the future STCF, more exciting topics will be studied and more attractive measurements will be performed, which will definitely benefit a lot the precise probe of the (anti-)hyperon-nucleus/nucleon interactions and provide constraints for the studies of the potential of strong interaction, the origin of color confinement, the unified model for baryon-baryon interactions, and the internal structure of neutron stars.

\bibliographystyle{ws-ijmpa}
\bibliography{references}

\end{document}